\documentclass[a4paper]{article}
\usepackage[AMS,theorem,useeps]{cjssac}
\usepackage{cleveref}
\usepackage{esvect}
\usepackage{url}

\Crefname{Th}{Theorem}{Theorems}
\Crefname{Cor}{Corollary}{Corollaries}
\Crefname{Pro}{Proposition}{Propositions}
\Crefname{Def}{Definition}{Definitions}

\renewcommand{\u}{\mathbf{u}}
\newcommand{\x}{\mathbf{x}}
\newcommand{\V}{\mathbf{V}}
\newcommand{\I}{\mathbf{I}}

\newcommand{\prem}{\mathrm{prem}}

\newcommand{\ideal}[1]{\langle{#1}\rangle}

\title{Computer-assisted proofs of ``Kariya's theorem'' with computer algebra}
\author{
	Ayane Ito
	\affil{University of Tsukuba} 
	\and 
	Takefumi Kasai
	\affil{University of Tsukuba} 
	\and
	Akira Terui
	\mail{terui@math.tsukuba.ac.jp}
	\affil{University of Tsukuba}
}

\begin{document}

    \begin{abstract}
        We demonstrate computer-assisted proofs of ``Kariya's theorem,''
        a theorem in elementary geometry, with computer algebra.
        In the proof of geometry theorem with computer algebra, 
        vertices of 
        geometric figures that are subjects for the proof are expressed as variables.
        The variables are
        classified into two classes: arbitrarily given points and 
        the points defined from the former points by constraints.
        We show proofs of Kariya's theorem with two formulations 
        according to two ways for giving the arbitrary points: 
        one is called ``vertex formulation,'' and the other is called
        ``incenter formulation,'' with two methods: 
        one is Gr\"obner basis computation, and the other is Wu's method.
        Furthermore, we show computer-assisted proofs of the property that
        the point so-called ``Kariya point'' is located on the hyperbola
        so-called ``Feuerbach's hyperbola'', with two formulations and  
        two methods.
    \end{abstract}

\section{Introduction}
\label{sec:intro}

This paper discusses computer-assisted proofs of ``Kariya's theorem,''
a theorem in elementary geometry with computer algebra.

In proving elementary geometry theorems with computer algebra,
the hypothesis and the conclusion are expressed as a system of polynomial equations 
and a polynomial equation, respectively.
In this formulation, 
variables appearing in the equations are divided into two sets: 
one consists of variables corresponding to arbitrarily given points, 
and the other one consists of variables corresponding to the points
derived from the points represented by the variables in the former set, with constraints.
The proof is demonstrated by showing 
that the algebraic variety defined by ``hypothesis''  equations
is included in the algebraic variety defined by the ``conclusion'' equation.
Generally, the ``computation'' of the proof is reduced to solving an ideal or a 
radical membership problem derived from ``hypothesis'' equations.
This computation is accomplished by Gr\"obner basis computation \cite{CLO2015}
or Wu's method \cite{cho1988}.

``Kariya's theorem'' \cite{kar1904} is a theorem in elementary geometry 
related to the incenter of a triangle. It was discovered at the end of 
the 19th century \cite{bostan2012} and is still a subject of study today
 (\cite{kiss-yiu2014}, \cite{yiu2015}).
In this paper, we show computational proof of the theorem with two different formulations: 
1) with the given points located on the vertices of the triangle (which is called 
``vertex formulation''), and 2) with the given point located in the vertices of the base and 
the incenter of the triangle
(which is called ``incenter formulation'').
For each formulation, we show the proof of Kariya's theorem and its corollary 
with the methods of Gr\"obner basis computation and Wu's method.
Furthermore, it is known that the point appearing in the assertion of Kariya's theorem
(so-called ``Kariya point'') is located on the rectangle hyperbola called 
``Feuerbach hyperbola.'' 
Therefore, we also show the computer-assisted proofs of this 
property with the two formulations above using Gr\"obner basis computation and Wu's method.
We note that, to the authors' knowledge, except for the proof of the corollary with incenter 
formulation (see \Cref{sec:cor:kariya-incenter}), 
the proofs shown in the present paper have never appeared in the literature.

The paper is organized as follows. 
In \Cref{sec:kariya}, Kariya's theorem and its formulations are shown.
In \Cref{sec:proof}, methods of computer-assisted proof of Kariya's theorem
with Gr\"obner basis computation and Wu's method are explained.
In \Cref{sec:experiments}, proofs of Kariya's theorem with 
Gr\"obner basis computation and Wu's method are demonstrated.
In \Cref{sec:feuerbach}, we show proofs of the property that the Kariya point is 
located on the Feuerbach hyperbola using the two methods with two different formulations.
Finally, we make concluding remarks in \Cref{sec:conclusion}.

\section{Kariya's theorem and its formulations}
\label{sec:kariya}

``Kariya's theorem'' is a theorem in elementary geometry describing a property related
to the incenter of a triangle. While several mathematicians have discovered
its proof (with generalizations of the original theorem)
during the end of the 19th century and the beginning 
of the 20th century~\cite{bostan2012}, The name of Kariya \cite{kar1904} has remained until today.

Kariya's theorem is as follows.
Note that a claim widely known as ``Kariya's theorem'' is a corollary of the following theorem 
(see Corollary~\ref{cor:kariya}).

\begin{Th}[Kariya's theorem \cite{kar1904}]
    \label{thm:kariya}
    In triangle $ABC$, let $O$ be the incenter of the triangle $ABC$, and let 
    $D'$, $E'$, and $F'$ be the points where the incenter circle touches 
    the sides $BC$, $CA$, and $AB$, respectively. For a real number $k$, 
    let $D$, $E$ and $F$ be the points on
    lines $OD'$, $OE'$ and $OF'$, respectively, satisfying that 
    $\vv{OD}=k\vv{OD'}$, $\vv{OE}=k\vv{OE'}$, $\vv{OF}=k\vv{OF'}$.
    Then, the lines $AD$, $BE$ and $CF$ are concurrent at a point $G$.
\end{Th}

The point $G$ in Theorem~\ref{thm:kariya} is called the ``Kariya point.''
In Theorem~\ref{thm:kariya}, by setting $k=1$, that is $D$, $E$ and $F$ coincides with
$D'$, $E'$ and $F'$, respectively, we obtain the following corollary.

\begin{Cor}
    \label{cor:kariya}
    In triangle $ABC$, let $D$, $E$ and $F$ be the points where the incenter circle touches 
    the sides $BC$, $CA$ and $AB$, respectively. Then, the lines 
    $AD$, $BE$ and $CF$ are concurrent at a point $G$.
\end{Cor}

In proving a theorem in elementary geometry with computer algebra, we give 
a coordinate system on the real plane (or space). 
Then, using the coordinates of the points appearing in the proof as variables, 
we express the relations on the geometric figures containing those points as 
algebraic relations of variables, which form polynomial equations.
As a result, we express the 
hypothesis and the conclusion as a system of polynomial equations and a polynomial
equation, respectively.
As a coordinate system, the cartesian coordinate system is widely used.

Some points appearing in the proof whose coordinates are to be expressed as variables 
are arbitrarily given, and others are derived from the arbitrarily given points. 
For the arbitrarily given points, let their coordinates be expressed as
$u_1,u_2,\dots,u_m$, and the coordinates of the points derived from 
the arbitrarily given points whose coordinate is expressed
with $u_i$ be expressed with $x_1,x_2,\dots,x_n$.
The variables $u_1,u_2,\dots,u_m$ are called ``free variables,'' and
the variables $x_1,x_2,\dots,x_n$ are called ``dependent variables.''

Properties of geometric figures are expressed as polynomial equations with respect to
the above variables. For convenience, tuples of variables are
denoted as $\u=(u_1,u_2,\dots,u_m)$ and $\x=(x_1,x_2,\dots,x_n)$. 
The number of polynomial equations expressing the hypothesis generally equals 
the number of dependent variables $m$. Thus, 
let the polynomial equations expressing
the hypothesis be
\[
    h_1(\u,\x)=0,\dots,h_n(\u,\x)=0,
\]
where $h_1(\u,\x),\dots,h_n(\u,\x)\in\mathbb{R}[\u,\x]$,
and the polynomial equation expressing the conclusion be $g(\u,\x)=0$,
where $g(\u,\x)\in\mathbb{R}[\u,\x]$.

With the discussion above, let us show formulations of Kariya's theorem 
into polynomial equations. 
We have two kinds of formulations according to setting the arbitrarily given points as 
\begin{enumerate}
    \item The vertices of triangle $ABC$, and
    \item The vertices of the base and the incenter of triangle $ABC$.
\end{enumerate}
The former is called ``vertex formulation,'' and the latter
is called ``the incenter formulation.''

\subsection{A formulation with setting arbitrarily given points as 
the vertices of a triangle (vertex formulation)}

In vertex formulation, without loss of generality, let $BC$ be the base of triangle $ABC$
with setting $A(u_1,u_2)$, $B(0,0)$, $C(0,1)$ where $u_1>0$ and $u_2>0$.
Formulations for Theorem~\ref{thm:kariya} and Corollary~\ref{cor:kariya} are shown as follows.

\subsubsection{The case of Theorem~\ref{thm:kariya}}
\label{sec:thm:kariya-vertex}

\begin{figure}
    \centering
    \includegraphics[scale=0.8]{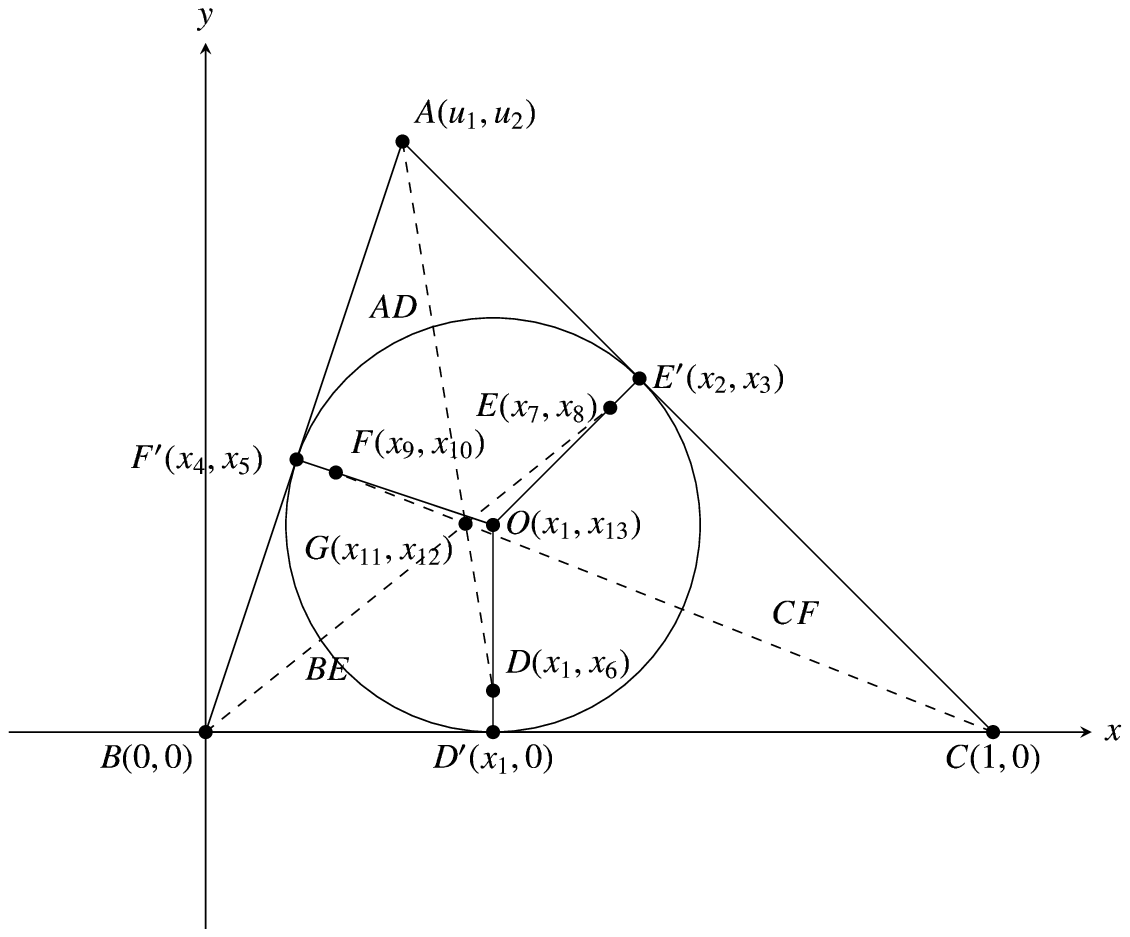}
    \caption{An example of Theorem~\ref{thm:kariya} with vertex formulation.
    See \Cref{sec:thm:kariya-vertex} for details.}
    \label{fig:kariya-theorem-vertex}
\end{figure}

In the formulation of Theorem~\ref{thm:kariya} (see \Cref{fig:kariya-theorem-vertex}),
let $O(x_1,x_{13})$ be the incenter and 
$D'(x_1,0)$, $E'(x_2,x_3)$ and $F'(x_4,x_5)$ be the points where the incenter circle 
touches the sides $BC$, $CA$ and $AB$, respectively.
Let $D(x_1,x_6)$, $E(x_7,x_8)$ and $F(x_9,x_{10})$ be the points on lines 
$OD'$, $OE'$ and $OF'$, respectively, satisfying that, for $k\in\mathbb{R}$,
$\vv{OD}=k\vv{OD'}$, $\vv{OE}=k\vv{OE'}$ and $\vv{OF}=k\vv{OF'}$.
Let $G(x_{11},x_{12})$ be the Kariya point.
Then, the hypothesis are expressed as $h_1,\dots,h_{13}$ in \Cref{eq:vertex-assumption-thm}, 
and the conclusion is expressed as $g_1$ in \Cref{eq:vertex-conclusion-thm}.
Note that, in each equation, an arrow ($\Longleftrightarrow$) followed by a comment 
on the right side shows the corresponding geometric condition.

\begin{equation}
    \label{eq:vertex-assumption-thm}
    \begin{split}
      h_1 &= x_1^2-x_4^2-x_5^2 \, \Longleftrightarrow \, BD'=BF',\\
      h_2 &= (1-x_1)^2-(1-x_2)^2-x_3^2 \, \Longleftrightarrow \, CD'=CE',\\
      h_3 &= (u_1-x_4)^2+(u_2-x_5)^2-(u_1-x_2)^2-(u_2-x_3)^2 \,
       \Longleftrightarrow \, AF'=AE', \\
      h_4 &= u_1x_5-u_2x_4 \, \Longleftrightarrow \, \text{$A$, $B$ and $F'$ are collinear,} \\
      h_5 &= u_2(1-x_2)-x_3(1-u_1) \, \Longleftrightarrow \, 
      \text{$A$, $E'$ and $C$ are collinear,} \\
      h_6 &= (x_2-x_1)^2+(x_3-x_{13})^2-x_{13}^2 \, \Longleftrightarrow \, OD' = OE', \\
      h_7 &= (x_6-x_{13})+kx_{13} \, \Longleftrightarrow \, \vv{OD}=k\vv{OD'}, \\
      h_8 &= (x_7-x_1)-k(x_2-x_1) \, \Longleftrightarrow \, \vv{OE}=k\vv{OE'}
      \quad\text{(with respect to the $x$ coordinate),}\\
      h_9 &= (x_8-x_{13})-k(x_3-x_{13}) \, \Longleftrightarrow \, \vv{OE}=k\vv{OE'}\quad
      \text{(with respect to the $y$ coordinate),} \\
      h_{10} &= (x_9-x_1)-k(x_4-x_1) \, \Longleftrightarrow \, \vv{OF}=k\vv{OF'}\quad
      \text{(with respect to the $x$ coordinate),} \\
      h_{11} &= (x_{10}-x_{13})-k(x_5-x_{13}) \, \Longleftrightarrow \, \vv{OF}=k\vv{OF'}\quad
      \text{(with respect to the $y$ coordinates),} \\
      h_{12} &= (u_2-x_6)(u_1-x_{11})-(u_1-x_1)(u_2-x_{12}) \, \Longleftrightarrow \, 
      \text{$A$, $G$ and $D$ are collinear,} \\
      h_{13} &= (1-x_{11})x_{10}-(1-x_9)x_{12} \, \Longleftrightarrow \, 
      \text{$C$, $G$ and $F$ are collinear.} 
    \end{split}
\end{equation}
\begin{equation}
\label{eq:vertex-conclusion-thm}
g = x_7x_{12}-x_8x_{11} \, \Longleftrightarrow \, \text{$B$, $G$ and $E$ are collinear.} 
\end{equation}

\subsubsection{The case of Corollary~\ref{cor:kariya}}
\label{sec:cor:kariya-vertex}

\begin{figure}
    \centering
    \includegraphics[scale=0.8]{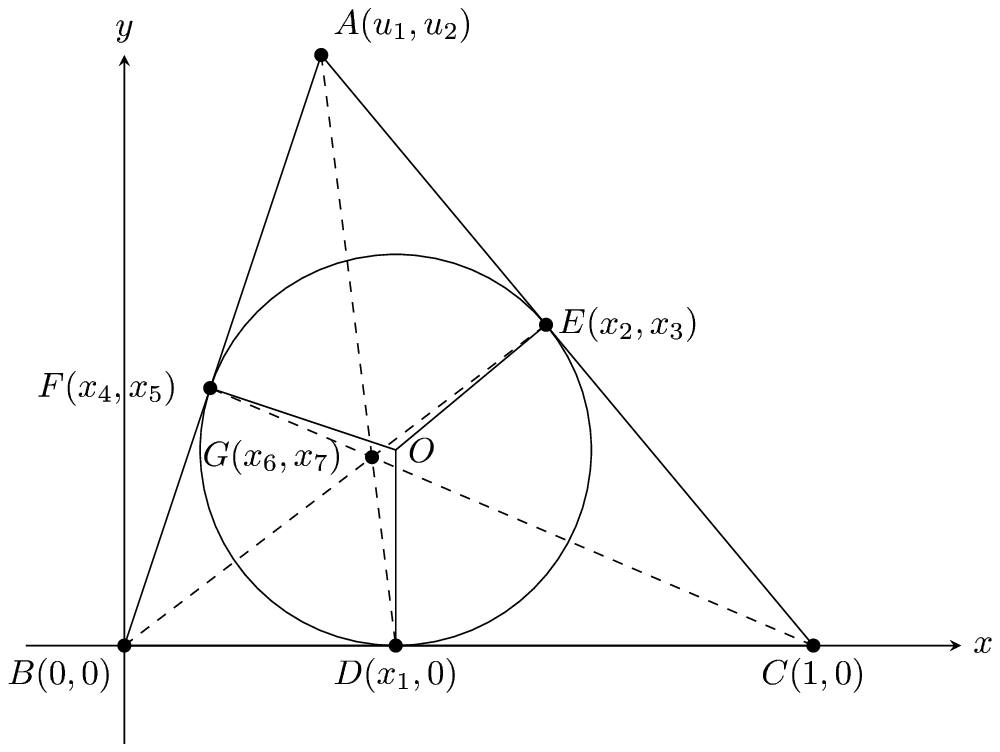}
    \caption{An example of Corollary~\ref{cor:kariya} with vertex formulation.
    See \Cref{sec:cor:kariya-vertex} for details.}
    \label{fig:kariya-corollary-vertex}
\end{figure}

In Corollary~\ref{cor:kariya} (see \Cref{fig:kariya-corollary-vertex}),
the points $D$, $E$ and $F$ coincides with 
$D'(x_1,0)$, $E'(x_2,x_3)$ and $F'(x_4,x_5)$, respectively, in 
Theorem~\ref{thm:kariya}.
Thus, in the formulation of Corollary~\ref{cor:kariya}, set
$D(x_1,0)$, $E(x_2,x_3)$, $F(x_4,x_5)$ and $G(x_6,x_7)$.
Then, the hypothesis are expressed as $h_1,\dots,h_{8}$ in \Cref{eq:vertex-assumption-cor}, 
and the conclusion is expressed as $g$ in \Cref{eq:vertex-conclusion-cor}.
\begin{equation}
    \label{eq:vertex-assumption-cor}
    \begin{split}
      h_1 &= x_1^2-x_4^2-x_5^2 \, \Longleftrightarrow \, BD=BF,\\
      h_2 &= (u_1-x_4)^2+(u_2-x_5)^2-(u_1-x_2)^2-(u_2-x_3)^2 \, \Longleftrightarrow \, AF=AE,\\
      h_3 &= (x_2-1)^2+x_3^2-(1-x_1)^2 \, \Longleftrightarrow \, CE=CD,\\
      h_4 &= u_1x_5-u_2x_4 \, \Longleftrightarrow \, \text{$A$, $F$ and $B$ are collinear,} \\
      h_5 &= u_2(1-x_2)-x_3(1-u_1) \, \Longleftrightarrow \,
      \text{$A$, $E$ and $C$ are collinear,} \\
      h_6 &= (u_1-x_1)x_7-(x_6-x_1)u_2 \, \Longleftrightarrow \, 
      \text{$A$, $G$ and $D$ are collinear,} \\
      h_7 &= x_2x_7-x_6x_3 \, \Longleftrightarrow \,  \text{$B$, $G$ and $E$ are collinear.}
    \end{split}
\end{equation}
\begin{equation}
    \label{eq:vertex-conclusion-cor}
    g = (x_4-1)x_7-(x_6-1)x_5 \, \Longleftrightarrow \, \text{$C$, $G$ and $F$ are collinear.} 
\end{equation}

\subsection{A formulation with setting arbitrarily given points as the vertices of the base 
and the incenter of the triangle (incenter formulation)}

In incenter formulation, without loss of generality, let $BC$ be the base of triangle $ABC$ 
with setting $B(0,0)$ and $C(1,0)$. 
Furthermore, let $O(u_1,u_2)$ be the incenter of triangle $ABC$.
Formulations for Theorem~\ref{thm:kariya} and Corollary~\ref{cor:kariya} are shown as follows.

\subsubsection{The case of Theorem~\ref{thm:kariya}}

\begin{figure}
    \centering
    \includegraphics[scale=0.8]{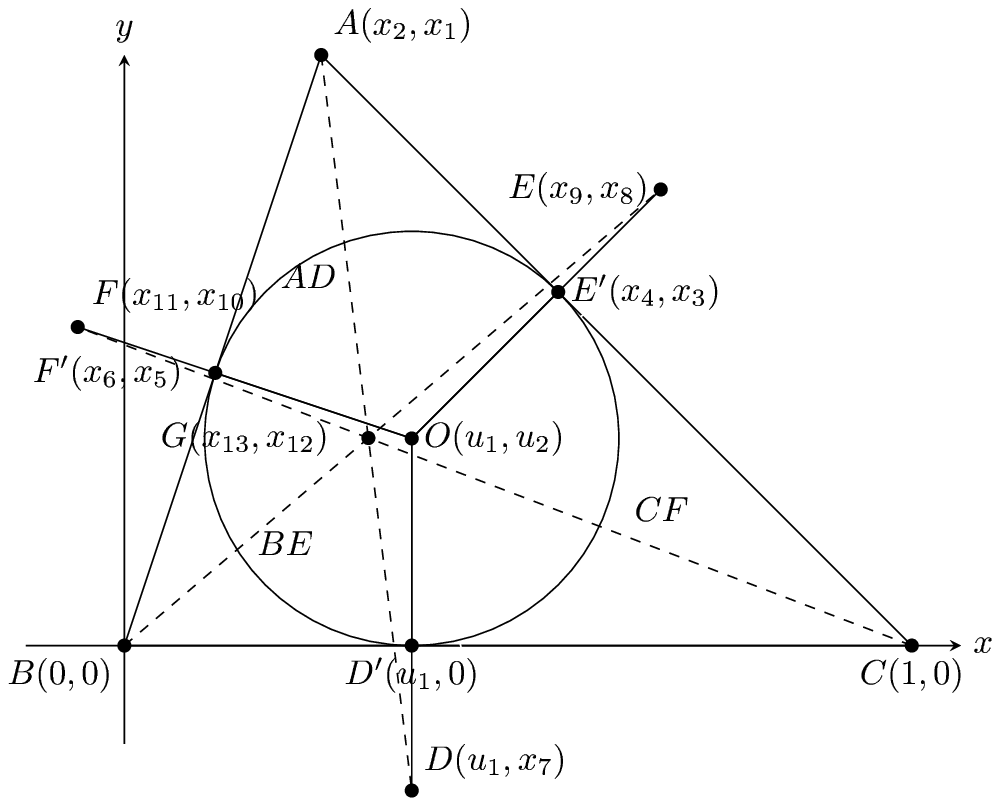}
    \caption{An example of Theorem~\ref{thm:kariya} with incenter formulation.
    Note that, in this figure, points $D$, $E$ and $F$ divides segments 
    $OD'$, $OE'$ and $OF'$ externally, respectively, while, in 
    \cref{fig:kariya-theorem-vertex}, points $D$, $E$ and $F$ divides segments 
    $OD'$, $OE'$ and $OF'$ internally, respectively.
    See \Cref{sec:thm:kariya-vertex} for details.}
    \label{fig:kariya-theorem-incenter}
\end{figure}

In the formulation of Theorem~\ref{thm:kariya} (see \Cref{fig:kariya-theorem-incenter}),
let $A(x_2,x_1)$ be the remaining vertex of the triangle, and let 
$D'(u_1,0)$, $E'(x_4,x_3)$ and $F'(x_6,x_5)$ be the points where the incenter circle touches 
the sides $BC$, $CA$ and $AB$, respectively.
Let $D(u_1,x_7)$, $E(x_9,x_8)$ and $F(x_{11},x_{10})$ be the points on the lines
$OD'$, $OE'$ and $OF'$, respectively, satisfying that, for $k\in\mathbb{R}$, 
$\vv{OD}=k\vv{OD'}$, $\vv{OE}=k\vv{OE'}$ and $\vv{OF}=k\vv{OF'}$.
Then, the hypothesis are expressed as $h_1,\dots,h_{13}$ in \cref{eq:incenter-assumption-thm},
and the conclusion is expressed as $g_1$ in \cref{eq:incenter-conclusion-thm}.

\begin{equation}
    \label{eq:incenter-assumption-thm}
    \begin{split}
      h_1 &= -(u_1-1)^2 x_1 + u_2^2 x_1 + 2 (u_1-1) u_2 (x_2-1)\, 
      \Longleftrightarrow \, 
      \tan{\angle BCO}=\tan{\angle OCA}, 
      \\
      h_2 &= x_1(-u_1^2 + u_2^2) + 2 u_1 u_2 x_2 \,
      \Longleftrightarrow \, \tan{\angle CBO} = \tan{\angle OBA},
      \\
      h_3 &= x_3(x_2-1)-x_1(x_4-1)  \, \Longleftrightarrow \, 
      \text{$A$, $E'$ and $C$ are collinear,}\\
      h_4 &= (x_2-1)(x_4-u_1)+x_1(x_3-u_2) \, \Longleftrightarrow \, 
      \text{$OE'$ and $CA$ intersect perpendicularly,} \\
      h_5 &= x_1x_6-x_2x_5  \, \Longleftrightarrow \, 
      \text{$A$, $F'$ and $B$ are collinear,}\\
      h_6 &= x_2(x_6-u_1)+x_1(x_5-u_2) \, \Longleftrightarrow \, 
      \text{$OF'$ and $BA$ intersect perpendicularly,} \\
      h_7 &= x_7+u_2(k-1) \, \Longleftrightarrow \, \vv{OD} = k \vv{OD'} \, 
      \text{(with respect to the $y$ coordinate),}   \\
      h_8 &= (x_9-u_1)-(k(x_4-u_1)) \, \Longleftrightarrow \, \vv{OE} = k \vv{OE'} \,
      \text{(with respect to the $x$ coordinate),}  \\
      h_9 &= (x_8-u_2)-(k(x_3-u_2)) \, \Longleftrightarrow \, \vv{OE} = k \vv{OE'} \,
      \text{(with respect to the $y$ coordinate),} \\
      h_{10} &= (x_{11}-u_1)-(k(x_6-u_1)) \, \Longleftrightarrow \, \vv{OF} = k \vv{OF'} \,
      \text{(with respect to the $x$ coordinate),} \\
      h_{11} &= (x_{10}-u_2)-(k(x_5-u_2)) \, \Longleftrightarrow \, \vv{OF} = k \vv{OF'} \,
      \text{(with respect to the $y$ coordinate),} \\
      h_{12} &= (x_{13}-u_1)(x_1-x_7)-(x_{12}-x_7)(x_2-u_1)  \, \Longleftrightarrow \,
      \text{$A$, $G$ and $D$ are collinear,}\\
      h_{13} &= x_{12}(x_{11}-1)-x_{10}(x_{13}-1) \, \Longleftrightarrow \, 
      \text{$C$, $G$ and $F$ are collinear.}\\
    \end{split}
\end{equation}
\begin{equation}
    \label{eq:incenter-conclusion-thm}
    g = x_8x_{13}-x_9x_{12}  \, \Longleftrightarrow \, 
    \text{$B$, $G$ and $E$ are collinear.}
\end{equation}

\subsubsection{The case of Corollary~\ref{cor:kariya}}
\label{sec:cor:kariya-incenter}

\begin{figure}
    \centering
    \includegraphics[scale=0.8]{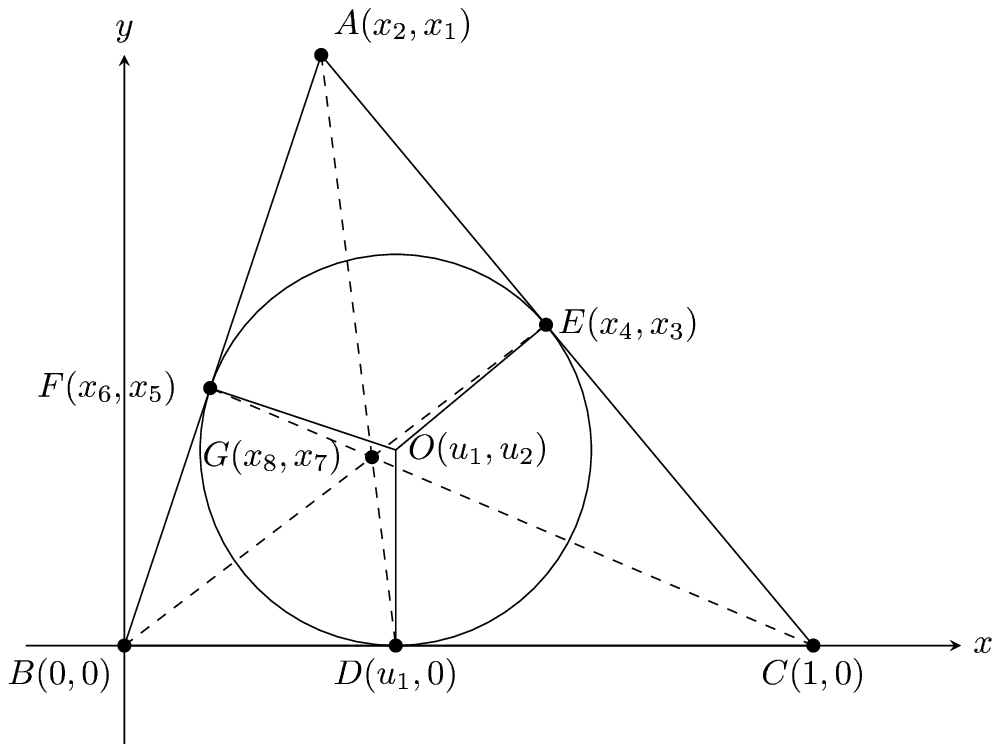}
    \caption{An example of Corollary~\ref{cor:kariya} with incenter formulation.
    See \Cref{sec:cor:kariya-incenter} for details.}
    \label{fig:kariya-corollary-incenter}
\end{figure}

In the formulation of Corollary~\ref{cor:kariya} (see \cref{fig:kariya-corollary-incenter})
As in \Cref{sec:cor:kariya-vertex}, set $D(u_1,0)$, $E(x_4,x_3)$ and $F(x_6,x_5)$.
Then, the hypothesis are expressed as $h_1,\dots,h_7$ in \cref{eq:incenter-assumption-cor},
and the conclusion is expressed as $g$ in \cref{eq:incenter-conclusion-cor}.
Note that this formulation has also been given in Chou \cite[Example 336]{cho1988}. 

\begin{equation}
    \label{eq:incenter-assumption-cor}
    \begin{split}
      h_1 &= u_2^2((1-x_4)^2+x_3^2)-(1-u_1)^2((u_1-x_4)^2+(u_2-x_3)^2) \, \Longleftrightarrow \, \tan{\angle BCO} = \tan{\angle OCA}, \\
      h_2 &= u_2^2(x_6^2+x_5^2)-u_1^2((u_1-x_6)^2+(u_2-x_5)^2) \, \Longleftrightarrow \,
       \tan{\angle CBO} = \tan{\angle OBA}, \\
      h_3 &= x_3(x_2-1)-x_1(x_4-1) \, \Longleftrightarrow \, 
      \text{$A$, $E$ and $C$ are collinear,} \\
      h_4 &= x_3(x_3-u_2)+(x_4-1)(x_4-u_1) \, \Longleftrightarrow \, 
      \text{$OE$ and $CE$ intersect perpendicularly,} \\
      h_5 &= x_6x_1-x_2x_5 \, \Longleftrightarrow \,  
      \text{$A$, $F$ and $B$ are collinear,}\\
      h_6 &= x_5(x_5-u_2)+x_6(x_6-u_1) \, \Longleftrightarrow \, 
      \text{$OF$ and $BF$ intersect perpendicularly,}\\
      h_7 &= x_8x_3-x_4x_7 \, \Longleftrightarrow \, 
      \text{$B$, $G$ and $E$ are collinear,}\\
      h_8 &= (x_8-u_1)x_1-x_7(x_2-u_1) \, \Longleftrightarrow \,  
      \text{$A$, $G$ and $D$ are collinear.}
    \end{split}
\end{equation}
\begin{equation}
    \label{eq:incenter-conclusion-cor}
    g = (x_8-1)x_5-x_7(x_6-1) \, \Longleftrightarrow \,  
    \text{$C$, $G$ and $F$ are collinear.}
\end{equation}

\section{Proofs of Kariya's theorem with Gr\"obner basis computation and Wu's method}
\label{sec:proof}

We review the fundamental theory of proving geometric theorem with computer algebra,
following Cox et al.\ \cite{CLO2015}.
Assume that the hypothesis are expressed as $h_1(\u,\x),\dots,h_n(\u,\x)$ and the conclusion is 
expressed as $g(\u,\x)$, where $\u$, $\x$, $h_1,\dots,h_n$ and $g$ are defined as the same as above.
Proving a theorem in elementary geometry can be reduced to 
showing that the real zeros of the equations a
$h_1(\u,\x)=\cdots=h_n(\u,\x)=0$ are also zeros of the equation $g(\u,\x)=0$.
This idea gives us a naive definition that one can deduce the conclusion.
In what follows, let $V=\V(h_1,\dots,h_n)\subset\mathbb{R}^{m+n}$ be 
the affine variety defined by $h_1,\dots,h_n$ and let $\I(V)$ be the ideal of $V$.

\begin{Def}[``follows strictly'']
    \label{def:follows-strictly}
    The conclusion $g$ follows \emph{strictly} from the hypothesis $h_1,\dots,h_n$ if 
    $g\in\I(V)$.
\end{Def}

\begin{Pro}
    \label{pro:follows-strictly}
    If $g\in\sqrt{\ideal{h_1,\dots,h_n}}$, then $g$ follows strictly from $h_1,\dots,h_n$.
\end{Pro}

From Definition~\ref{def:follows-strictly} and Proposition~\ref{pro:follows-strictly}, a naive proof of the theorem 
is reduced to solving the radical membership problem.
However, this condition seems too strict because the converse of Proposition~\ref{pro:follows-strictly} may not be true.
In such a case, there may exist a polynomial $h(\u)$ in the hypothesis containing only independent variables $\u$, and $h(\bar{\u})=0$ for $\bar{\u}=(\bar{u}_1,\dots,\bar{u}_m)\in V$,
which means a degenerate case of the configuration of geometric figures \cite{CLO2015}.
To avoid such degenerate cases, we handle a subvariety of $V$ satisfying that
for the points in which 
a defining polynomial with only independent variables is always nonzero, as
in the following definition.

\begin{Def}[Algebraically independent]
    Let $W\subseteq R^{m+n}$ be an irreducible affine variety with the coordinates
    $u_1,\dots,u_m,x_1,\dots,x_n$. 
    The variables $u_1,\dots,u_m$ are algebraically independent on $W$ if
    there exist no nonzero polynomial with variables $u_1,\dots,u_m$ 
    that has zeros in $W$, that is, $u_1,\dots,u_m$ satisfy that
    $\I(W)\cap\mathbb{R}[u_1,\dots,u_m]=\{0\}$.
\end{Def}

Then, we accept non-degenerate cases for the geometric proving with the following
definition.

\begin{Def}[``follows generically'']
    \label{def:follows-generically}
    The conclusion $g$ follows \emph{generically} from the hypothesis $h_1,\dots,h_n$
    if 
    \[
        g\in\I(V')\subseteq R[u_1,\dots,u_m,x_1,\dots,x_n],  
    \]
    where $V'=W_1\cup\cdots\cup W_p\subset\mathbb{R}^{m+n}$ satisfying that,
    for $i=1,\dots,p$, $W_i$ is irreducible and 
    $u_1,\dots,u_m$ are algebraically independent on $W_i$.
\end{Def}

Initially, for deriving a proof with Definition~\ref{def:follows-generically}, one needs to compute
irreducible components of $V=\V(h_1,\dots,h_n)$. 
Fortunately, we have the following proposition.

\begin{Pro}
    \label{pro:follows-generically}
    Let $H=\ideal{h_1,\dots,h_n}$.
    If there exists a nonzero polynomial $c(u_1,\dots,u_m)\in\mathbb{R}[u_1,\dots,u_m]$ 
    satisfying that $c\cdot g\in\sqrt{H}$, then
    the conclusion $g$ follows generically from the hypothesis $h_1,\dots,h_n$.
\end{Pro}

Note that, if $g\in H$, then $g$ and $H$ satisfy Proposition~\ref{pro:follows-generically}.

\subsection{Computing a proof with Gr\"obner basis computation}

This section explains computing a proof with Gr\"obner basis computation \cite{CLO2015}.
Proposition~\ref{pro:follows-generically} tells us that computing a proof is reduced to solving 
the radical membership problem.
We have the following corollary.

\begin{Cor}
    \label{cor:follows-generically}
    Under the conditions of Proposition~\ref{pro:follows-generically}, the following are equivalent.
    \begin{enumerate}
        \item There exists a nonzero polynomial $c(u_1,\dots,u_m)\in\mathbb{R}[u_1,\dots,u_m]$ 
        satisfying that $c\cdot g\in\sqrt{H}$.
        \item Let $\tilde{H}$ be an ideal in $\mathbb{R}(u_1,\dots,u_m)[x_1,\dots,x_n]$ 
        generated by $h_1,\dots,h_n$. Then, we have $g\in\sqrt{\tilde{H}}$.
        \item The reduced Gr\"obner basis of an ideal
        $\ideal{h_1,\dots,h_n,1-yg}\subseteq\mathbb{R}(u_1,\dots,u_m)[x_1,\dots,x_n,y]$
        is equal to $\{1\}$.
    \end{enumerate}
\end{Cor}

Corollary~\ref{cor:follows-generically} tells us that computing the proof is reduced to
either solving the ideal membership problem $g\in H$ or computing the reduced 
Gr\"ober basis of the ideal $\ideal{h_1,\dots,h_n,1-yg}$.

\subsection{Computing a proof with Wu's method}

In this section, we explain computing proof with Wu's method.
Note that the method of computation presented here is an elementary version 
of Wu's method \cite{CLO2015}, and a complete version of it can be found 
in other literature (for example, see Chou \cite{cho1988}).
In Wu's method, we first ``triangulate'' the polynomials corresponding to
the hypothesis by pseudo-divisions. Then, we repeat pseudo-divisions on 
the polynomial corresponding to the conclusion by the triangulated polynomials
to show that the conclusion follows from the hypothesis.

\begin{Pro}[Pseudo-division \cite{CLO2015}]
    \label{pro:pseudodiv}
    Let $f,g \in k[x_{1}, \ldots, x_{n}, y]$ be polynomials expressed as
       \begin{equation}
        \label{eq:fg}
          f = c_{p}y^{p}+ \ldots +c_{1}y+c_{0}, \quad
          g = d_{m}y^{m}+ \ldots +d_{1}y+d_{0},
       \end{equation}
    where $c_{i},d_{i}\in k[x_{1},\ldots,x_{n}]$ with $m\leq p$ and $d_{m}\neq 0$.
    Then, there exist polynomials $q,r\in k[x_{1},\ldots,x_{n}]$ satisfying the following conditions.
    \begin{enumerate}
        \item $r=0$, or $\deg_yr<m$ and there exists a nonnegative integer $s$ satisfying
        $d_m^sf=qg+r$.
        \item $r\in\ideal{f,g}$ in the ring $k[x_{1},\ldots,x_{n},y]$.
    \end{enumerate}
\end{Pro}

In Proposition~\ref{pro:pseudodiv}, polynomials $q$ and $r$ are called a \emph{pseudoquotient} and 
a \emph{pseudoremainder}, respectively, of $f$ on pseudo-division by $g$ with respect to $y$.
The pseudoremainder $r$ is denoted by $\prem(f,g,y)$.

In the algorithm of pseudo-division, $d_m^s$ is chosen such that the division is executed 
in the polynomial ring $k[x_{1},\ldots,x_{n}]$. 
Furthermore, in place of $d_m^s$, $d_m/\gcd(d_m,c_p,\dots,c_0)$ can be used 
for avoiding the growth of degrees of coefficient polynomials \cite{Shiraishi-Kai-Noda2002}.

In ``triangulation'' of the hypothesis polynomials $h_{1},\ldots,h_{n}\in k[\u,\x]$,
pseudo-divisions with respect to variables $x_n,x_{n-1}\ldots,x_1$ is executed repeatedly 
for reducing to a ``triangulated'' system of polynomials
\begin{equation}    
    \label{eq:ascending-chain}
    \begin{split}
        f_{1}(x_{1}),
        f_{2}(x_{1}, x_{2}),
        \dots,
        f_{n}(x_{1}, \ldots , x_{n}). 
    \end{split}
\end{equation}  
The order of variables used for computing $f_1,\dots,f_n$ is denoted by 
\[
    x_n\succ x_{n-1}\succ\cdots\succ x_1, 
\]
and the set of polynomials in \cref{eq:ascending-chain} is called an 
\emph{ascending chain}.

\begin{Def}[Irreducible ascending chain]
    An ascending chain of polynomials in \cref{eq:ascending-chain} is called 
    \emph{irreducible} if, for $i=1,\dots,n$, $f_i$ is irreducible in the polynomial ring 
    $k(u_1,\dots,u_m)[x_i,\dots,x_i]/\ideal{f_1,\dots,f_{i-1}}$.
\end{Def}

Then, for the conclusion polynomial $g\in k[\u,\x]$, pseudo-division by 
the polynomials in the ascending chain \cref{eq:ascending-chain} is repeated for computing 
polynomials $R_{n-1},\dots,R_{n-0}$ as 
\begin{equation}
    \label{eq:prem-chain}
    R_{n-1}=\prem(g,f_n,x_n),\, R_{n-2}=\prem(R_{n-1},f_{n-1},x_{n-1}),
    \dots, R_{0}=\prem(R_{1},f_{1},x_{1}),
\end{equation}
and $R_0$ is denoted by $\prem(g,f_1,\dots,f_n)$.
We have the following proposition.

\begin{Pro}
    \label{pro:wu-proof}
    Let $\{f_1,\dots,f_n\}$ be an ascending chain derived from the 
    hypothesis polynomials $h_1,\dots,h_n$ expressed as in \cref{eq:ascending-chain}, and
    let $g$ be the conclusion polynomial.
    Then, the following are equivalent.
    \begin{enumerate}
        \item $\prem(g,f_1,\dots,f_n)=0$.
        \item There exists a nonzero polynomial $c(\u)\in \mathbb{R}[\u]$ satisfying that 
        $c\cdot g\in\ideal{f_1,\dots,f_n}$.
    \end{enumerate} $\prem(g,h_1,\dots,h_n)$
\end{Pro}

In Proposition~\ref{pro:wu-proof}, note that we have $\ideal{f_1,\dots,f_n}\subset H$,
where $H$ is defined as in 
Proposition~\ref{pro:follows-generically}, since $f_i\in H$.
Thus, Corollary~\ref{cor:follows-generically} and Proposition~\ref{pro:wu-proof} tells us that, 
if we have $\prem(g,f_1,\dots,f_n)=0$, then the conclusion $g$ follows generically from
the hypothesis $h_1,\dots,h_n$.

\section{Experiments}
\label{sec:experiments}

We have implemented an elementary version of Wu's method on the 
Computer Algebra System (CAS) Risa/Asir \cite{nor2003}, and have computed 
proofs of Theorem~\ref{thm:kariya} and Corollary~\ref{cor:kariya} with the 
Gr\"obner basis computation and Wu's method using 
the vertex and the incenter formulations \cite{ter-ito-kas-kariya2023}.
The test was conducted in the following environment: 
Intel Xeon Silver 4210 at 2.20 GHz, RAM 256 GB, Linux 5.4.0 (SMP), Asir Version 20210326.

\subsection{Computing proofs with the Gr\"obner basis computation}

This section separately explains computing proofs with the Gr\"obner basis computation 
for the vertex and the incenter formulations.

\subsubsection{Computing proofs using the vertex formulation}
\label{sec:kariya-groebner-vertex-experiment}

In computing the proof of Theorem~\ref{thm:kariya}, for the hypothesis polynomials $h_1,\dots,h_{13}$ in 
\cref{eq:vertex-assumption-thm},
we have computed a Gro\"bner basis $G_1$ of the ideal $I=\ideal{h_1,\dots,h_{13}}$ with respect to 
the degree reverse lexicographic (DegRevLex) ordering with the variable order given as
\begin{equation}
    \label{eq:groebner-proof-theorem-vertex-variable-order}    
    x_6\succ x_7\succ x_8\succ x_9\succ x_{10}\succ x_{11}\succ x_{12}\succ x_1\succ x_2\succ x_3\succ x_4\succ x_5\succ x_{13}.
\end{equation}
Then, for the conclusion polynomial $g$ in \cref{eq:vertex-conclusion-thm},
we have verified that $g\in H=\ideal{h_1,\dots,h_{13}}$ by showing that the normal form of 
$g$ with respect to $G_1$ is equal to $0$.

In computing the proof of Corollary~\ref{cor:kariya}, for the hypothesis polynomials $h_1,\dots,h_{7}$ in 
\cref{eq:vertex-assumption-cor}, 
we have computed a Gro\"bner basis $G_2$ of the ideal $I=\ideal{h_1,\dots,h_{7}}$ with respect to 
the DegRevLex ordering with the variable order given as
$x_7\succ x_6\succ x_5\succ x_4\succ x_3\succ x_2\succ x_1$.
Then, for the conclusion polynomial $g$ in \cref{eq:vertex-conclusion-cor},
we have verified that $g\in H=\ideal{h_1,\dots,h_{7}}$ by showing that the normal form of 
$g$ with respect to $G_2$ is equal to $0$.

\subsubsection{Computing proofs using the incenter formulation}
\label{sec:kariya-groebner-incenter-experiment}

In computing the proof of Theorem~\ref{thm:kariya}, for the hypothesis polynomials $h_1,\dots,h_{13}$ in 
\cref{eq:incenter-assumption-thm}, 
we have computed a Gro\"bner basis $G_3$ of the ideal $I=\ideal{h_1,\dots,h_{13}}$ with respect to
the DegRevLex ordering with the variable order given as
$x_{13}\succ x_{12}\succ x_{11}\succ x_{10}\succ x_9\succ x_8\succ x_7\succ x_6\succ x_5\succ x_4\succ x_3\succ x_2\succ x_1$.
Then, for the conclusion polynomial $g$ in \cref{eq:incenter-conclusion-thm}, 
we have verified that $g\in H=\ideal{h_1,\dots,h_{13}}$ by showing that the normal form of 
$g$ with respect to $G_3$ is equal to $0$.

In computing the proof of Corollary~\ref{cor:kariya}, for the hypothesis polynomials $h_1,\dots,h_{7}$ in 
\cref{eq:incenter-assumption-cor}, 
we have computed a Gro\"bner basis $G_4$ of the ideal $I=\ideal{h_1,\dots,h_{7}}$ with respect to 
the DegRevLex ordering with the variable order given as
$x_8\succ x_7\succ x_6\succ x_5\succ x_4\succ x_3\succ x_2\succ x_1$.
We have computed that the normal form of the conclusion polynomial $g$ in 
\cref{eq:incenter-conclusion-cor} with respect $G_4$ is not equal to zero, 
and the reduced Gr\"obner basis of the ideal 
$\ideal{h_1,\dots,h_{7},1-yg}$ is not equal to $\{1\}$.
Then, by adding a constraint that $u_1\ne 0$ (with a new variable $y_2$), 
we have computed that the reduced Gr\"obner basis of the ideal 
\begin{equation}
    \label{eq:incenter-cor-groebner-radical}
    \ideal{h_1,\dots,h_{7},1-yg,1-y_2u_1}  
\end{equation}
equals $\{1\}$.

\subsubsection{Computing time and the variable ordering}
\label{sec:groebner-time-order}

\Cref{tab:computing-time-groebner} shows the computing time of computing the proofs in this section.
For each formulation, the table shows the computing times of the Gr\"obner basis and the normal form.
The exception is the proof of Corollary~\ref{cor:kariya} using the incenter formulation: 
in this case, the table shows only the computing time of the reduced Gr\"obner basis of the ideal in \cref{eq:incenter-cor-groebner-radical}.
In the tables below, computing time with the letter $\dag$, $\ddag$ and $\dag\dag$ 
denote the average of repeatedly measured data for 10, 100, and 1000 times, respectively.

In each computation of the proof, variable ordering is defined as follows.
For the proof of Theorem~\ref{thm:kariya} using the vertex formulation, 
in the hypothesis polynomials in \cref{eq:vertex-assumption-thm}, 
the variables $x_6,x_7,x_8,x_9,x_{10},x_{11},x_{12}$ appear in 
the terms of total degree $1$. 
Thus, we have defined the variable ordering as in 
\cref{eq:groebner-proof-theorem-vertex-variable-order}
for reducing the terms in $x_6,x_7,x_8,x_9,x_{10},x_{11},x_{12}$ first.
In the other cases, since the computing time of 
the Gr\"obner basis, as well as the normal form, was sufficiently small,
we have defined the variable ordering as 
$x_{13}\succ x_{11}\succ\cdots\succ x_1$ for the proof of Theorem~\ref{thm:kariya} and
$x_{8}\succ x_{7}\succ\cdots\succ x_1$ for the proof of Corollary~\ref{cor:kariya}.

\begin{table}[t]
    \centering
    \caption{Computing time of the proofs with Gr\"obner basis computation. 
    Note that computing time with the letter $\dag$, and $\dag\dag$ 
    denote the average of repeatedly measured data for 10 and 1000 times, respectively.
    See \Cref{sec:groebner-time-order} for details.}
    \begin{tabular}{cccc}
      \hline
      Formulation & Theorem & \multicolumn{2}{c}{Computing time (sec.)} \\
      & & Gr\"obner basis & The normal form \\
      \hline
      Vertex formulation & Theorem~\ref{thm:kariya} & $762.0^{\dag}$ & $0.4194^{\dag}$ \\
      & Corollary~\ref{cor:kariya} & $0.524^{\dag}$ & $0.02152^{\dag}$ \\
      \hline
      Incenter formulation & Theorem~\ref{thm:kariya} & $0.05179^{\dag}$ & $0.007314^{\dag}$ \\
      & Corollary~\ref{cor:kariya} & $0.002069^{\dag\dag}$ &  N/A \\
      \hline
    \end{tabular}
    \label{tab:computing-time-groebner}
\end{table}

\subsection{Computing proofs with Wu's method}

In this section, we explain computing proofs with Wu's method 
separately for the vertex and the incenter formulations.

\subsubsection{Computing proofs using the vertex formulation}

In computing the proof of Theorem~\ref{thm:kariya}, we have set the order of variables as
\begin{equation}
    \label{eq:wu-proof-theorem-vertex-variable-order}
    x_8\succ x_7\succ x_{12}\succ x_{11}\succ x_{10}\succ x_9\succ x_6\succ x_{13}\succ x_5\succ x_4\succ x_3\succ x_2\succ x_1,
\end{equation}
and, for $h_1,\dots,h_{13}$ in \cref{eq:vertex-assumption-thm}, computed an ascending chain as 
\[
    f_{1,1}(x_1),f_{1,2}(x_1,x_2),\dots,f_{1,13}(x_8,x_7,x_{12},\dots,x_2,x_1).
\]
Then, for $g$ in \cref{eq:vertex-conclusion-thm}, 
we have computed $\prem(g,f_{1,1},\dots,f_{1,13})=0$.

In computing the proof of Corollary~\ref{cor:kariya}, we have set the order of variables as
$x_7\succ x_6\succ x_5\succ x_4\succ x_3\succ x_2\succ x_1$, 
and, for $h_1,\dots,h_7$ in \cref{eq:vertex-assumption-cor}, computed an ascending chain as
\[
    f_{2,1}(x_1),f_{2,2}(x_1,x_2),\dots,f_{2,7}(x_7,\dots,x_2,x_1).
\]
Then, for $g$ in \cref{eq:vertex-conclusion-cor}, 
we have computed $\prem(g,f_{2,1},\dots,f_{2,7})=0$.

\subsubsection{Computing proofs using the incenter formulation}
\label{sec:wu-incenter}

In computing the proof of Theorem~\ref{thm:kariya}, we have set the order of variables as
$x_{13}\succ x_{12}\succ x_{11}\succ x_{10}\succ x_9\succ x_8\succ x_7\succ x_6\succ x_5\succ x_4\succ x_3\succ x_2\succ x_1$,
and, for $h_1,\dots,h_{13}$ in \cref{eq:incenter-assumption-thm}, computed an ascending chain as 
\[
    f_{3,1}(x_1),f_{3,2}(x_1,x_2),\dots,f_{3,13}(x_{13},x_{12},\dots,x_2,x_1).
\]
Then, for $g$ in \cref{eq:incenter-conclusion-thm}, 
we have computed $\prem(g,f_{3,1},\dots,f_{3,13})=0$.

In computing the proof of Corollary~\ref{cor:kariya}, we have set the order of variables as
$x_8\succ x_7\succ x_6\succ x_5\succ x_4\succ x_3\succ x_2\succ x_1$,
and, for $h_1,\dots,h_8$ in \cref{eq:incenter-assumption-cor}, computed an ascending chain as
\[
    f_{4,1}(x_1),f_{4,2}(x_1,x_2),\dots,f_{4,8}(x_{8},\dots,x_2,x_1).
\]
Then, for $g$ in \cref{eq:incenter-conclusion-cor}, 
we have computed $\prem(g,f_{4,1},\dots,f_{4,8})=0$.

\subsubsection{Computing time and the variable ordering}
\label{sec:wu-time-order}

\Cref{tab:computing-time-wu} shows the computing time of computing this section's proofs.
For each formulation, the table shows the total computing time for calculating the ascending chain 
and for computing repeated pseudo-divisons of the conclusion polynomial as in \cref{eq:prem-chain}.

In each computation of the proof, the order of variables has been defined as follows. 
For the proof of Theorem~\ref{thm:kariya} using vertex formulation,
the conclusion polynomial $g$ in \cref{eq:vertex-conclusion-thm} has variables
$x_7,x_8,x_{11},x_{12}$.
Furthermore, in the hypothesis polynomials in \cref{eq:vertex-assumption-thm}, 
there are polynomials with $x_7$ and $x_8$ of degree 1, respectively.
Thus, we have aimed to eliminate $x_7$ and $x_8$ from $g$ first, then
$x_{11}$ and $x_{12}$ from pseudoremainders.
After that, since there exist hypothesis polynomials in \cref{eq:vertex-assumption-thm}
which have terms in $x_{11}$, $x_{10}$, $x_9$, $x_6$ and $x_{13}$ of degree 1, 
we have aimed to eliminate these variables. 
As a result, we have defined the order of variables as in 
\cref{eq:wu-proof-theorem-vertex-variable-order}.
In the other cases, since the computing time was sufficiently short, 
we have defined the order of variables as 
$x_{13}\succ x_{12}\succ\cdots\succ x_1$ for the proof of Theorem~\ref{thm:kariya} and 
$x_{8}\succ x_{7}\succ\cdots\succ x_1$ for the proof of Corollary~\ref{cor:kariya}.

\begin{table}[t]
    \centering
    \caption{Computing time of the proofs with Wu's method.
    Note that computing time with the letter $\dag$, and $\dag\dag$ 
    denote the average of repeatedly measured data for 10 and 1000 times, respectively.
    See \Cref{sec:wu-time-order} for details.}
    \begin{tabular}{ccc}
      \hline
      Formulation & Theorem & Computing time (sec.)\\
      \hline
      Vertex formulation & Theorem~\ref{thm:kariya} & $2.952^{\dag}$ \\
      & Corollary~\ref{cor:kariya} & $0.005355^{\dag\dag}$ \\
      \hline
      Incenter formulation & Theorem~\ref{thm:kariya} & $0.00124^{\dag\dag}$ \\
      & Corollary~\ref{cor:kariya} & $0.01187^{\dag\dag}$ \\
      \hline
    \end{tabular}
    \label{tab:computing-time-wu}
\end{table}

\section{Computation on the Feuerbach hyperbola}
\label{sec:feuerbach}

For a given triangle, the Feuerbach hyperbola is a rectangular hyperbola
centered at the point of contact of the nine-point circle and the incircle 
and passing the triangle's vertices.
Furthermore, it is known that for changing the value of $k$ in Theorem~\ref{thm:kariya}, 
the Kariya point is located on the Feuerbach hyperbola \cite{kiss-yiu2014}.
This section, shows this property with the Gr\"obner basis computation 
and Wu's method using vertex and incenter formulations.

For $a\in\mathbb{R}$, a rectangular hyperbola whose focus is located at  
$(\pm\sqrt{2}\,a,0)$ is expressed as
\begin{equation}
    \label{eq:hyperbola-standard}
    x^2-y^2=a^2.
\end{equation}
By translating the center to $(p_x,p_y)$ and rotating $\theta$ counterclockwise,
where $p_x,p_y,\theta\in\mathbb{R}$, the hypothesis in \cref{eq:hyperbola-standard} 
becomes as 
\begin{equation}
    \label{eq:hyperbola-general}
    (cx+sy-p_x)^2-(-sx+cy-p_y)^2=a^2,
\end{equation}
where $c=\cos\theta$, $s=\sin\theta$.

The proofs are computed as follows. 
From \cref{eq:hyperbola-general},
let $g'=(cx+sy-p_x)^2-(-sx+cy-p_y)^2-a^2$ with $x$ and $y$ are replaced with 
appropriate variables. 
After computing the Gr\"obner basis or the ascending set from the hypothesis
polynomials in Theorem~\ref{thm:kariya} or Corollary~\ref{cor:kariya}, 
add the constraint $c^2+s^2-1$ to the Gr\"obner basis or the ascending set. 
If the result of the reduction of $g'$ by the set of polynomials is equal to $0$, 
we see that the Kariya point is located on the Feuerbach hyperbola.

\subsection{Computing the proof with Gr\"obner basis computation}

Gr\"obner basis computation has been used for computing the proofs as follows.

In computing the proof with the vertex formulation,
for the Gr\"obner basis $G_1$ computed in \Cref{sec:kariya-groebner-vertex-experiment},
let $\bar{G}_1=\{c^2+s^2-1\}\cup G_1$. 
Using \cref{eq:hyperbola-general}, let
\[
    g'=(cx_{11}+sx_{12}-p_x)^2-(-sx_{11}+cx_{12}-p_y)^2-a^2 \Longleftrightarrow \,  
    \text{$G(x_{11},x_{12})$ is located on the hyperbola},
\]
and we have computed that the normal form of $g'$ with respect to $\bar{G}_1$ 
is equal to $0$ to show $g'\in\ideal{\bar{G}_1}$.

In computing the proof with the incenter formulation,
for the Gr\"obner basis $G_3$ computed in \Cref{sec:kariya-groebner-incenter-experiment},
let $\bar{G}_3=\{c^2+s^2-1\}\cup G_3$. 
Using \cref{eq:hyperbola-general}, let
\[
    g'=(cx_{13}+sx_{12}-p_x)^2-(-sx_{13}+cx_{12}-p_y)^2-a^2 \Longleftrightarrow \,  
    \text{$G(x_{13},x_{12})$ is located on the hyperbola},
\]
and we have computed that the normal form of $g'$ with respect to $\bar{G}_3$ 
is equal to $0$ to show $g'\in\ideal{\bar{G}_3}$.

\subsection{Computing the proof with Wu's method}

Wu's method has been used for computing the proofs as follows.

In computing the proof with the vertex formulation,
for the hypothesis polynomials $h_1,\dots, h_9,$ $h_{12}, h_{13}$ in 
\cref{eq:vertex-assumption-thm}, 
we have computed an ascending chain 
\[
    f_{5,1}(x_5),f_{5,2}(x_5,x_{13}),\dots,f_{5,11}(x_{5},\dots,x_8,x_{10}),
\]
with respect to the order of variables given as
\begin{equation}
    \label{eq:feuerbach-wu-vartex-var-order}
    x_{10}\succ x_8\succ x_7\succ x_6\succ x_{11}\succ x_{12}\succ x_1\succ x_2\succ x_3\succ x_4\succ x_{13}\succ x_5.
\end{equation}
Then, let 
\[
    g' = (c x_{11}+s x_{12}-p_x)^2-(-s x_{11}+c x_{12}-p_y)^2-a^2 \Longleftrightarrow \,  
    \text{$G(x_{11},x_{12})$ is located on  the hyperbola},
\]
and we have computed $\prem(g,f_{5,1},\dots,f_{5,11})=0$.

In computing the proof with the incenter formulation,
for a hyperbola in \cref{eq:hyperbola-standard},
translate the center to $(p_x,p_y)$ and rotate $\theta$ counterclockwise, 
and let $c=\cos2\theta$, $s=\sin2\theta$.
Let the set of hypothesis polynomials consists of 
$h_1,\dots,h_9,h_{12},g$ in \cref{eq:incenter-assumption-thm}, and
\begin{equation}
    \label{eq:feuerbach-wu-incenter-new-polys}
    \begin{split}
      h_{14} &= c^2+s^2-1 \quad \text{(a constraint on $\sin2\theta$ and $\cos2\theta$),} \\
      h_{15} &= c-2p_xc-2p_ys \, \Longleftrightarrow \, 
      \text{$B(0,0)$ and $C(1,0)$ are located on the hyperbola,}\\
      h_{16} &= x_2^2c-2x_2p_xc+2x_1x_2s-2x_2p_ys-2p_xx_1s-x_1^2c+2x_1p_yc\\
        &\quad \Longleftrightarrow \, \text{$B(0,0)$ and $A(x_2,x_1)$ are located on the hyperbola,} \\
      h_{17} &= u_1^2c-2u_1p_xc+2u_1u_2s-2u_1p_ys-2p_xu_2s-u_2^2c+2u_2p_yc\\
        &\quad \Longleftrightarrow \, \text{$B(0,0)$ and $O(u1,u2)$ are located on the hyperbola.}
    \end{split}
\end{equation}
For the set of the hypothesis polynomials, we have computed an ascending chain
\[
    f_{6,1}(x_1),f_{6,2}(x_1,x_2),\dots,f_{6,15}(x_{1},\dots,p_y,p_x),
\]
with respect to the order of variables as
\begin{equation}
    \label{eq:feuerbach-wu-incenter-var-order}
    p_x\succ p_y\succ c\succ s\succ x_{13}\succ x_{12}\succ x_9\succ x_8\succ x_7\succ x_6\succ x_5\succ x_4\succ x_3\succ x_2\succ x_1.   
\end{equation}
Then, let
\[
    \begin{split}
      g' &= x_{13}^2c-2x_{13}p_xc+2x_{13}x_{12}s-2x_{13}p_ys-2p_xx_{12}s-x_{12}^2c+2x_{12}p_yc\\
      &\quad \Longleftrightarrow \, \text{$B(0,0)$ and $G(x_{13},x_{12})$
        are located on the hyperbola},
    \end{split}
\]
and we have computed $\prem(g_2,f_{6,1},\dots,f_{6,15})=0$.

Note that the derivation of $h_{15},h_{16},h_{17},g_2$ will be explained in the Appendix.

\subsection{Computing time and the variable ordering} 
\label{sec:feuerbach-time-order}

\Cref{tab:computing-time-feuerbach} shows the computing time of the proofs 
in this section, with Gr\"obner basis computation and Wu's method,
using the vertex and the incenter formulations.

The Ordering of variables is defined as follows.
In the Gr\"obner basis computation, the order of variables used 
for the proof of Theorem~\ref{thm:kariya} are used (see \Cref{sec:groebner-time-order}).

In Wu's method with the vertex formulation, the order of variables are given as 
in \cref{eq:feuerbach-wu-vartex-var-order} by the following reason.
When we compute the ascending chain, we first eliminate $x_{10}$ because the number of terms
in $x_{10}$ which appear in $h_1,\dots, h_9, h_{12}, h_{13}$ is the smallest 
among the variables which appear in $h_1,\dots, h_9, h_{12}, h_{13}$. 
Next, we eliminate $x_{8}$ because the number of terms in $x_{8}$ which 
appear in the input polynomials is the smallest among the variables
which appear in the input polynomials.
By repeating the procedure, we eliminate the variable in which the number of terms 
appearing in the polynomials is the smallest in each step in computing the 
ascending chain.

In Wu's method with the incenter formulation, the order of variables is given as
in \cref{eq:feuerbach-wu-incenter-var-order} by the following reason.
In computing the ascending chain, we first eliminate newly added variables 
$p_x$, $p_y$, $c$, and $s$ in this order, then eliminate the rest of the variables
with the same ordering as the computation for the proof of Theorem~\ref{thm:kariya} 
(see \Cref{sec:wu-incenter}).

\begin{table}[t]
    \centering
    \caption{Computing time of the proofs on Feuerbach hyperbola.
    Note that computing time with the letter $\dag$, $\ddag$ and $\dag\dag$ denote the average of repeatedly measured data for 10, 100, and 1000 times, respectively.
    See \Cref{sec:feuerbach-time-order} for details.}
    \begin{tabular}{ccc}
      \hline
      Computing method & Formulation & Computing time (sec.)\\
      \hline
      Gr\"obner basis  & Vertex formulation & $0.01147^{\dag\dag}$ \\
      & Incenter formulation & $0.002255^{\dag\dag}$ \\
      \hline
      Wu's method & Vertex formulation & $0.1639^{\ddag}$ \\
      & Incenter formulation & $2.162^{\dag}$ \\
      \hline
    \end{tabular}
    \label{tab:computing-time-feuerbach}
\end{table}

\section{Concluding remarks}
\label{sec:conclusion}

In this paper, we have demonstrated computational proofs of Kariya's theorem and its corollary
with the Gr\"obner basis computation and Wu's method using the vertex and the incenter 
formulations. 
Furthermore, we have demonstrated computational proofs of the property that the Kariya point is 
similarly located on the Feuerbach hyperbola.

Computing time (see \Cref{tab:computing-time-groebner,tab:computing-time-wu})
 suggests that the incenter formulation is more suitable for
efficient computation for the proof of Theorem~\ref{thm:kariya}.
For the proof of Corollary~\ref{cor:kariya}, while using the incenter formulation made 
computation more efficient with Gr\"obner basis computation,
using the vertex formulation made  computation more efficient with Wu's method,
thus formulation used for better efficiency was different depending on the methods.

Future research topics on computer-assisted proof of Kariya's theorem with computer algebra
include the following.
\begin{enumerate}
    \item In Gr\"obner basis computation and Wu's method for the proofs, other variable
    orderings than those used in the present paper may speed up the computation.
    \item Setting the incenter to the origin may speed up 
    the computation using the incenter formulation.
    \item While Kariya's theorem uses the incenter, the theorem may hold for the excenter(s).
    \item With Gr\"obner basis computation, the formula of Feuerbach hyperbola may be derived 
    from the hypothesis polynomials.
    \item Although the cartesian coordinate system was used in this paper, other coordinate systems may speed up the computation. (Note that Coand\c{a} et al.\ \cite{coanda2013} use 
    the barycentric coordinate system for deriving Kariya's theorem from their theorem in 
    a more general form.)
\end{enumerate} 

\acknowledgements{The research in this paper has been initiated as an undergraduate research 
project in the College of Mathematics, School of Science and Engineering, University of Tsukuba.
The authors thank Nanako Ishii and Gaku Kuriyama for collaborating with the authors 
during the project.}


\section*{Appendix: Derivation of $h_{15},h_{16},h_{17}$ and $g_2$ in 
\Cref{eq:feuerbach-wu-incenter-new-polys}} 

In the appendix, we show derivation of $h_{15},h_{16},h_{17}$ and $g_2$ in 
\cref{eq:feuerbach-wu-incenter-new-polys}.

Before deriving the formulas, we show a transform on the hyperbola in 
\cref{eq:hyperbola-standard}.
By rotating the hyperbola in \cref{eq:hyperbola-standard} 
$\theta$ counterclockwise, we have
\[
    (x\cos\theta+y\sin\theta)^2-(-x\sin\theta+y\cos\theta)^2=a^2.
\]
Expanding the left-hand side and collecting the terms with respect to $x$ and $y$ 
is expressed as
\[
    x^2(\cos^2\theta-\sin^2\theta)-2xy(2\cos\theta\sin\theta)-y^2(\cos^2\theta-\sin^2\theta)=a^2.
\]   
By applying the double angle formula, we have
\[
    x^2c-2xys-y^2c=a^2,
\]
where $c=\cos2\theta$, $s=\sin2\theta$.
By translating the origin to $(p_x,p_y)$, we have
\begin{equation}
    \label{eq:feuerbach-hyperbola-appendix-standard}
    (x-p_x)^2c-2(x-p_x)(y-p_y)s-(y-p_y)^2c=a^2.
\end{equation}

Now, $h_{15}$ is derived as follows.
In \Cref{fig:kariya-theorem-incenter},
since the hyperbola in \cref{eq:feuerbach-hyperbola-appendix-standard} passes through
$B(0,0)$, we have 
\begin{equation}
    \label{eq:feuerbach-hyperbola-appendix-b}
    p_x^2c-2p_xp_ys-p_y^2c=a^2. 
\end{equation}
Furthermore, since the same hyperbola passes through $C(1,0)$, we have
\begin{equation}
    \label{eq:feuerbach-hyperbola-appendix-c}
    (1-p_x)^2c-2(1-p_x)p_ys-p_y^2c=a^2.
\end{equation}
By equating the left-hand-sides of 
\cref{eq:feuerbach-hyperbola-appendix-b,eq:feuerbach-hyperbola-appendix-c},
$h_{15}$ is derived.

Next, $h_{16}$ is derived as follows.
In \Cref{fig:kariya-theorem-incenter},
since the hyperbola in \cref{eq:feuerbach-hyperbola-appendix-standard} passes through
$A(x_2,x_1)$, we have
\begin{equation}
    \label{eq:feuerbach-hyperbola-appendix-a}
    (x_2-p_x)^2c-2(x_2-p_x)(x_1-p_y)s-(x_1-p_y)^2c=a^2.
\end{equation}
By equating the left-hand-sides of 
\cref{eq:feuerbach-hyperbola-appendix-b,eq:feuerbach-hyperbola-appendix-a},
$h_{16}$ is derived.

Next, $h_{17}$ is derived as follows.
In \Cref{fig:kariya-theorem-incenter},
since the hyperbola in \cref{eq:feuerbach-hyperbola-appendix-standard} passes through
$O(u_1,u_2)$, we have
\begin{equation}
    \label{eq:feuerbach-hyperbola-appendix-O}
    (u_1-p_x)^2c-2(u_1-p_x)(u_2-p_y)s-(u_2-p_y)^2c=a^2.
\end{equation}
By equating the left-hand-sides of 
\cref{eq:feuerbach-hyperbola-appendix-b,eq:feuerbach-hyperbola-appendix-O},
$h_{17}$ is derived.

Finally, $g_{2}$ is derived as follows.
In \Cref{fig:kariya-theorem-incenter},
since the hyperbola in \cref{eq:feuerbach-hyperbola-appendix-standard} passes through
$G(x_{13},x_{12})$, we have
\begin{equation}
    \label{eq:feuerbach-hyperbola-appendix-g}
    (x_{13}-p_x)^2c-2(x_{13}-p_x)(x_{12}-p_y)s-(x_{12}-p_y)^2c=a^2.
\end{equation}
By equating the left-hand-sides of 
\cref{eq:feuerbach-hyperbola-appendix-b,eq:feuerbach-hyperbola-appendix-g},
$g_{2}$ is derived.

\end{document}